\documentclass[universe,article,accept,pdftex,moreauthors]{Definitions/mdpi}
\usepackage{macros}
\newcommand{\xadded}[1]{#1}
\newcommand{\xdeleted}[1]{}

\firstpage{1} 
\pubvolume{1}
\issuenum{1}
\articlenumber{0}
\pubyear{2022}
\copyrightyear{2022}
\externaleditor{Academic Editor: Firstname Lastname %MDPI: please add the academic editors.
}
\datereceived{} 
\dateaccepted{} 
\datepublished{} 
\hreflink{https://doi.org/} % If needed use \linebreak
\Title{Observe Gamma-Rays %MDPI: please check if the ``Gamma'' should be ``γ'' ($\gamma$).
 and Neutrinos Associated with Ultra-High Energy Cosmic Rays}

\TitleCitation{Observe Gamma-rays and Neutrinos Associated with Ultra-High Energy Cosmic Rays}

\Author{Qinyuan Zhang %MDPI: Please carefully check the accuracy of names and affiliations.
 $^{1}$, Xishui Tian $^{1}$ and Zhuo Li $^{1,2,}$*}

\AuthorNames{Qinyuan Zhang, Xishui Tian and Zhuo Li}

\AuthorCitation{Zhang, Q.; Tian, X.; Li, Z.}
\address{%
$^{1}$ \quad Department of Astronomy, School of Physics, Peking University, Beijing 100871, China; zhangqy@stu.pku.edu.cn (Q.Z.); tianxs@stu.pku.edu.cn (X.T.)\\
$^{2}$ \quad Kavli Institute for Astronomy and Astrophysics, Peking University, Beijing 100871, China}

\corres{Correspondence: zhuo.li@pku.edu.cn}

\abstract{IceCube measures a diffuse neutrino flux comparable to the Waxman-Bahcall bound, which suggests the possibility that the ultra-high energy cosmic rays (UHECRs) have a common origin with diffuse high energy neutrinos. We propose high energy gamma-ray and/or neutrino observations toward the arrival directions of UHECRs to search for the sources and test this possibility. We calculate the detection probability of gamma-ray/neutrino sources, and find that the average probability per UHECR of >10 EeV is $\sim$10\% if the sensitivity of the gamma-ray or neutrino telescope is \mbox{$\sim$10$^{-12}$ erg cm$^{-2}$s$^{-1}$} and the source number density is $\sim$10$^{-5}$ Mpc$^{-3}$. Future gamma-ray and neutrino observations toward UHECRs, e.g., by LHAASO-WCDA, CTA, IceCube/Gen2, are encouraged to constrain the density of UHECR sources or even identify the sources of UHECRs.}


\begin{document}

%%%%%%%%%%%%%%%%%%%%%%%%%%%%%%%%%%%%%%%%%%

\section{Introduction}

The origin of ultra-high energy cosmic rays ($\gtrsim$10$^{19}$ eV; UHECRs) is still an open question {\citep{UHECR_Nagano,UHECR_review}}. Because~cosmic rays are charged particles, their arrival directions will be affected by magnetic field during propagation, which makes it difficult to locate the UHECR sources by their arrival directions directly. The~sample size is another difficulty of searching sources by UHECR observations. The~cosmic ray flux above the cosmic ray spectral ``ankle'' is small and there is a cutoff feature at the highest energies {\citep{HiRes_cutoff,Auger_cutoff}}. The~ankle might be attributed to the transition from Galactic to extra-galactic components {\citep{galactic_ex_transition,Katz_energyrate}}, or~the ``dip'' due to the electron pair production in cosmic ray propagation \citep{dipmodel}. The~cutoff may be caused by interactions of UHECRs with the cosmic background radiation, known as Greisen-Zatsepin-Kuzmin (GZK) cutoff \citep{GZK1,GZK2}, however the maximum energy of particle acceleration {\citep{Hillas1984}} in the sources may also lead to such cutoff. These features of cosmic ray energy spectrum indicate that UHECRs may come from extra-galactic~sources.

Cosmic rays may produce high energy gamma-rays and neutrinos via pion production processes, when cosmic rays interact with background photons or matter. This makes it possible to build \xadded{a} connection between UHECRs and gamma-rays \xadded{or} neutrinos. Neutral particles are not deflected by magnetic fields, pointing back to the sources. Moreover, neutrinos hardly interact with other matter, which makes them a special kind of astrophysical messengers {\citep{Halzen2002,Waxman_neuastronomy}}.

High energy astrophysical neutrinos are discovered by the IceCube neutrino\linebreak \mbox{detector \citep{IceCube_1st_PeV,IceCube_evidence}}, opening a window to high energy neutrino astronomy. The~IceCube measured neutrino flux \citep{IceCube_flux} is comparable to the Waxman-Bahcall bound \citep{WBbound1,WBbound2}, \mbox{$\sim$2 $\times$ 10$^8$ GeV} cm$^{-2}$ s$^{-1}$ sr$^{-1}$, which is an upper bound for the extra-galactic neutrino flux  derived from the observed UHECR flux and assuming all the UHECR energy \xadded{are} converted to pions. Such coincidence may suggest that the UHECRs and high energy neutrinos are produced in the same or related processes \citep{UHECR_WBbound}. The~cosmic ray particles resulting in secondary neutrinos lose energy efficiently in the sources or the environment by pion production, thus \xadded{the neutrino flux from charged pion decay} can be comparable to the Waxman-Bahcall~bound.

In this work, we assume \xadded{a} common origin of UHECRs and the high energy neutrinos, and~suggest an observation strategy that search for gamma-ray and neutrino sources at UHECR directions. We calculate the probability of the detection of gamma-ray or neutrino sources in the UHECR~directions.

%%%%%%%%%%%%%%%%%%%%%%%%%%%%%%%%%%%%%%%%%%
\section{Physical Picture and~Assumption}
We adopt the following physical picture for the emission of gamma-ray and neutrinos associated with UHECR sources, and~some assumptions in the calculation of the gamma-ray and neutrino~emission.
\subsection{Uhecr Sources as Gamma-Ray and Neutrino~Sources}

We assume the UHECRs are dominantly protons, although~there is debate of proton versus heavy nucleus composition. In~this case, the~spectral cut-off feature at highest energy is well explained by the GZK effect, and~the observed spectral index above $10^{19}$eV is consistent with explained by shock acceleration \citep{shock_acc}.

%The coincidence of IceCube neutrino flux and Waxman-Bahcall bound implies a common origin for UHECRs and high energy astrophysical neutrino. We accept the following picture to explain the coincidence. 
We assume UHECRs and the high energy neutrino producing cosmic rays are accelerated by the same sources. The~UHECRs escape from the sources and their environment, e.g.,~the host galaxy, promptly due to their large rigidity, while the cosmic rays with lower energy are trapped in the host galaxy. These trapped cosmic rays interact with matter, e.g.,~the interstellar medium, in~the host galaxy and lose most of their energy in pion production process, then neutrinos and gamma-rays can be produced by pion decay. In~this situation, we naturally expect a diffuse neutrino flux from UHECR sources roughly equal to the Waxman-Bahcall bound, and~IceCube diffuse neutrino flux is mainly contributed by these UHECR associated~neutrinos.

The host galaxies emit both high energy neutrinos and gamma-rays. Neutrinos come from charged pion decay, $\pi^{\pm} \rightarrow \mu + \nu_{\mu}$, and~the following muon decay, $\mu \rightarrow e + \nu_{e} + \nu_{\mu}$. Neutral pions $\pi^{0}$’s are also produced in pion production process, and~they quickly decay to high energy gamma-rays, $\pi^{0} \rightarrow \gamma \gamma$. In~this picture, high energy neutrinos and gamma-rays are well connected in arrival directions and flux with UHECRs, if~UHECRs do not change their directions in propagation, which is the case as shown~below.

\subsection{\xadded{Spatial} and Temporal Association of UHECRs with Gamma-Ray/Neutrino~Signals}

%To show the feasibility of finding correlation between UHECR event and gamma-ray or neutrino source, we consider the their connection in space and time. 
Consider the propagation of protons with energy $E$ in the magnetic field with strength $B$ and correlation length $\lambda$. After~travel a distance $d$, the~typical angle deflection of the proton is $\theta $$\sim$$ (d/\lambda)^{\frac{1}{2}} (\lambda/R_{L})$, where $R_{L} = E/eB$ is Lamour radius of the proton \citep{Waxman_UHECRGRB}, i.e.,~$\theta $$\sim$$ 1.8^{\circ}(E/10^{20}{\rm \, eV})^{-1}(d/100{\rm \, Mpc})^{1/2}(B/1 {\rm \, nG})(\lambda/1 {\rm \, Mpc})^{1/2}$. %Gamma-ray photons and neutrinos are neutral particles, they are not affected by the magnetic field and can trace the source directly. 
It is expected to observe gamma-ray and neutrino point sources toward the UHECR directions within few degree~separation.

Due to the deflection, the~arrival time of UHECRs will be delayed relative to a photon emitted in the same time by a typical delay time, $\Delta t $$\sim$$ \theta^{2}d/c$$\sim$$ 10^{5} {\rm \, yr}$ for $E $$\sim$$ 10^{20} \rm \, eV$ and a distance $d $$\sim$$ 100\rm \, Mpc$, The arrival time will spread in a timescale similar to $\Delta t$, if~the UHECRs are produced by explosive events rather than steady~sources.

We assume the IceCube measured diffuse neutrino flux can be accounted for by UHECR sources and their environment. Thus we expect to observe gamma-rays and neutrinos from the UHECR directions in the same time for steady sources---however, even for explosive sources of UHECRs, this is also very likely. For~the low energy cosmic rays to be trapped in the source environment, say the source’s host galaxy, and~lose energy significantly in \xadded{pion} production, it may be required that the energy loss time scale is larger than the escape time scale of cosmic rays, thus the gamma-ray and neutrino emission duration will be the energy loss time of cosmic rays, $\Delta t_{\pi} $$\sim$$ 1/c\,n_{ISM} \sigma_{pp} $$\sim$$ 10^{7}  \rm \, yr$, given pp cross section $\sigma_{pp} $$\sim$$ 100 \rm \, mb$ for $\rm PeV$ protons, medium density $n_{ISM} $$\sim$$ 1 \rm \, cm^{-3}$, much larger than the UHECR spreading time. We expect to find a persistent source of gamma-rays or neutrinos around the direction of a detected UHECR.
%%%%%%%%%%%%%%%%%%%%%%%%%%%%%%%%%%%%%%%%%%
\section{Method and~Formula}\label{sec3}

The way to estimate the probability of finding correlated gamma-ray and neutrino sources is shown in this~section.

\subsection{Horizon for Gamma-Ray and Neutrino~Sources}\label{sec3.1}
%Basing on our assumptions, every UHECR source is also high energy gamma-ray and neutrino source, but not all sources can be confirmed by the detection. 
Both gamma-ray and neutrino source detection are limited by the sensitivity of telescopes. In~addition, high energy gamma-rays from extra-galactic sources are attenuated due to the absorption by the extra-galactic background light (EBL), $\gamma + \gamma_{EBL} \rightarrow e^{+} + e^{-}$. So only sources within some ``horizon'' can be detected. %Considering such limitation, we will make some estimation about the detection probability of gamma-ray source and neutrino source related to~UHECR.

The horizon is determined by the typical luminosity of the source. %We assume that there is one kind of source dominate the flux of UHECRs and IceCube neutrinos, and~we can take the average luminosity as the typical value for the single source. 
For gamma-ray sources, the~typical specific gamma-ray luminosity, $dL_\gamma/dE_\gamma$, of~a single source related to UHECRs can be derived via $\xi_{z} \frac{c}{4 \pi} t_{H} n_{s} E_{\gamma} \frac{\mathrm{d}L_{\gamma}}{\mathrm{d}E_{\gamma}} = E^{2}_{\gamma} \phi_{\gamma}$, i.e.,
\begin{equation}
     \frac{\mathrm{d}L_{\gamma}}{\mathrm{d}E_{\gamma}} =\xadded{ \frac{4\pi E_{\gamma} \phi_{\gamma}}{\xi_{z} c t_{H} n_{s}}}
\end{equation}
where $n_s$ is the source number density in the local universe, $t_H$ is the Hubble timescale of the universe, $\xi_z$ is a coefficient accounting for the redshift evolution of the UHECR production rate density\xadded{ and we take $\xi_z$ $\simeq$ 3 for redshift evolution following the star formation rate~\citep{WBbound1}}, and~$E_{\gamma}^2 \phi_{\gamma}$ is the total flux of gamma-ray from UHECR sources when EBL absorption is~neglected. 

The neutrino and gamma-ray are both from pion decays. We assume in $pp$ interactions, %MDPI: Please unify the format ``pp'' in the whole text (italic or normal).
 pion number ratio is $\pi^0$:$\pi^+$:$\pi^-$$\approx1$:$1$:$1$. Each gamma-ray gains $1/2$ of energy of a $\pi^{0}$, while each neutrino gains $1/4$ of energy of a $\pi^{\pm}$. Then we have $E_{\gamma}^2\phi_{\gamma} =% E_{\gamma}^2 \mathrm{d}N/\mathrm{d}E_{\gamma} = 
(1/2) E_{\pi}^2\phi_{\pi}$, and~$E_{\nu}^2\phi_{\nu} = %E_{\nu}^2 \mathrm{d}N/\mathrm{d}E_{\nu} = 
(1/4) E_{\pi}^2\phi_{\pi}$. So high energy gamma-ray flux (without EBL absorption) from UHECR sources can be scaled with neutrino flux as
\begin{equation}
    E^{2}_{\gamma} \phi_{\gamma} = 2E_{\nu}^2 \phi_{\nu}
\end{equation}
with $E_\gamma=2E_\nu$. From~the latest results of  IceCube observations, we have \mbox{$ \phi_{\nu} = 1.44 \times 10^{-18}$} $(E_{\nu}/100\, \rm TeV)^{-2.28} \, GeV^{-1} \, cm^{-2} \, s^{-1} \, sr^{-1}$ \citep{IceCube_icrc_flux} for single flavor (we have assumed equal flavor ratio for the three neutrino flavors).

Considering the EBL absorption for high energy gamma-rays, the~maximum redshift $z_{\max}$ of the gamma-ray sources that can be detected is determined by equation
\begin{equation}
    \int_{> E_{\gamma}} \mathrm{d}E_{\gamma}^{\prime} \frac{\mathrm{d}L_{\gamma}}{\mathrm{d}E_{\gamma}^{\prime}} e^{-\tau_{EBL}(E_{\gamma}^{\prime},z_{\max,\gamma})} = 4\pi d_L^2(z_{\max,\gamma}) \overline{f}_{th}(>E_{\gamma}).
\end{equation}

Here $\tau_{EBL}$ is the optical depth of high energy gamma-rays due to EBL absorption, for~which we adopt the observational results from \citep{EBL_depth}. $\overline{f}_{th}(>E_{\gamma})$ is the integral sensitivity of gamma-ray telescope at gamma-ray energy above $E_\gamma$ and $d_L$ is the luminosity distance  (we adopt the standard $\Lambda$CDM cosmological model). The~sensitivity $\overline{f}_{th}(>E_{\gamma})$ and the source number density $n_{s}$ are two parameters that decide the  maximum~redshift. 

The horizon for neutrino source searching can be derived similarly. Due to the very weak interactions with the background matter, the~universe is transparent to neutrinos, thus the neutrino horizon can be solved out via
\begin{equation}
    \int_{> E_{\nu}} \mathrm{d}E_{\nu}^{\prime} \frac{\mathrm{d}L_{\nu}}{\mathrm{d}E_{\nu}^{\prime}}  = 4\pi d_L^2(z_{\max,\nu}) \overline{f}_{th}(>E_{\nu})
\end{equation}
where $\overline{f}_{th}(>E_{\nu})$ is the neutrino telescope sensitivity, and~the typical specific neutrino luminosity $dL_\nu/dE_\nu$ for a single source can be obtained by
\begin{equation}
    \frac{\mathrm{d}L_{\nu}}{\mathrm{d}E_{\nu}} = \xadded{\frac{4\pi E_{\nu} \phi_{\nu}}{\xi_{z} c t_{H} n_{s}}}.
\end{equation}

\subsection{Detection~Probability}

We propose to use gamma-ray and/or neutrino telescopes to observe the directions of the detected UHECRs, and~estimate the average probability to detect a gamma-ray and/or neutrino source for per~UHECR. 

For an UHECR detector array with the effective area $A_{eff}$  and the observational time $\Delta T$, the~total number of UHECRs that are detected above some energy $E$ is the integration of the UHECR production rate over the universe history,
\begin{equation}
    N(>E) = \Delta T \int_0^{\infty} \mathrm{d}z \int_{E(1+z)}^{\infty} \mathrm{d}E^{\prime} \frac{\mathrm{d}V}{\mathrm{d}z} Q(E^{\prime},z) e^{-\tau_{GZK}(E^{\prime},z)} \frac{A_{eff}(E^{\prime}/(1+z)) (1+z)}{4\pi \mathrm{d}_L^2(z)},
\end{equation}
where $Q(E,z) \propto (E/E_{0})^{-\alpha} (1+z)^{m}$ is the specific production rate density of UHECRs as function of cosmic ray energy $E$ and at redshift $z$, d$V/$d$z$ is the comoving volume per unit redshift, and~$\tau_{GZK}$ accounts for the GZK effect (see below).

To account for the GZK effect, consider an ``optical depth'' $\tau_{GZK} = d(z)/l_{GZK}$, where $d(z)$ is the comoving distance and $l_{GZK}$ is the energy loss length due to GZK effect.\linebreak  Following~\cite{GZK_parameterization}, we take $l_{GZK}(E,z) = c (1+z)^{3} t((1+z)E)$, with~$t$ being the energy loss time\linebreak of UHECRs in the Cosmic Microwave Background, which is parameterized as\linebreak $t^{-1}(E) = t_{0,ep}^{-1} e^{-E_{c,ep}/E} + t_{0,\pi}^{-1} e^{-E_{c,\pi}/E}$, with~$E_{c,ep} = 2.7 \times 10^{18} \rm \, eV$, $t_{0,ep} = 3.4 \times 10^{9} \rm \, yr$, $E_{c,\pi} = 3.2 \times 10^{20} \rm \, eV$, and~$t_{0,\pi} = 2.2 \times 10^{7} \rm \, yr$.

Given the telescope horizon, the~number of detected UHECRs that are originated from sources within the horizon, i.e.,~within the maximum redshift, is % has been set for telescope sensitivity and source number density. UHECR sources in this horizon are detectable for gamma-ray or neutrino telescope, and~only UHECRs from such sources can have correlation with gamma-ray or neutrino sources. We use the ratio,

\vspace{-9pt}

\begin{adjustwidth}{-\extralength}{0cm}
%\centering %% If there is a figure in wide page, please release command
 \centering
\begin{equation}
    N(>E,z_{\max,\gamma/\nu}) = \Delta T \int_{0}^{z_{\max,\gamma/\nu}}\mathrm{d}z \int_{E(1+z)}^{\infty} \mathrm{d}E^{\prime} \frac{\mathrm{d}V}{\mathrm{d}z} Q(E^{\prime},z) e^{-\tau_{GZK}(E^{\prime},z)} \frac{A_{eff}(E^{\prime}/(1+z)) (1+z)}{4\pi d_L^2(z)}.
\end{equation}
\end{adjustwidth}

The ratio between the above two gives the average probability of finding a source from each detected UHECR with energy above $E$,
\begin{equation}
    P_{\gamma/\nu}(>E) = \frac{N(>E,z_{\max,\gamma/\nu})}{N(>E)}.
\end{equation}

\section{Results}\label{sec4}

\begin{figure}[H]

    \includegraphics[width=0.98\textwidth]{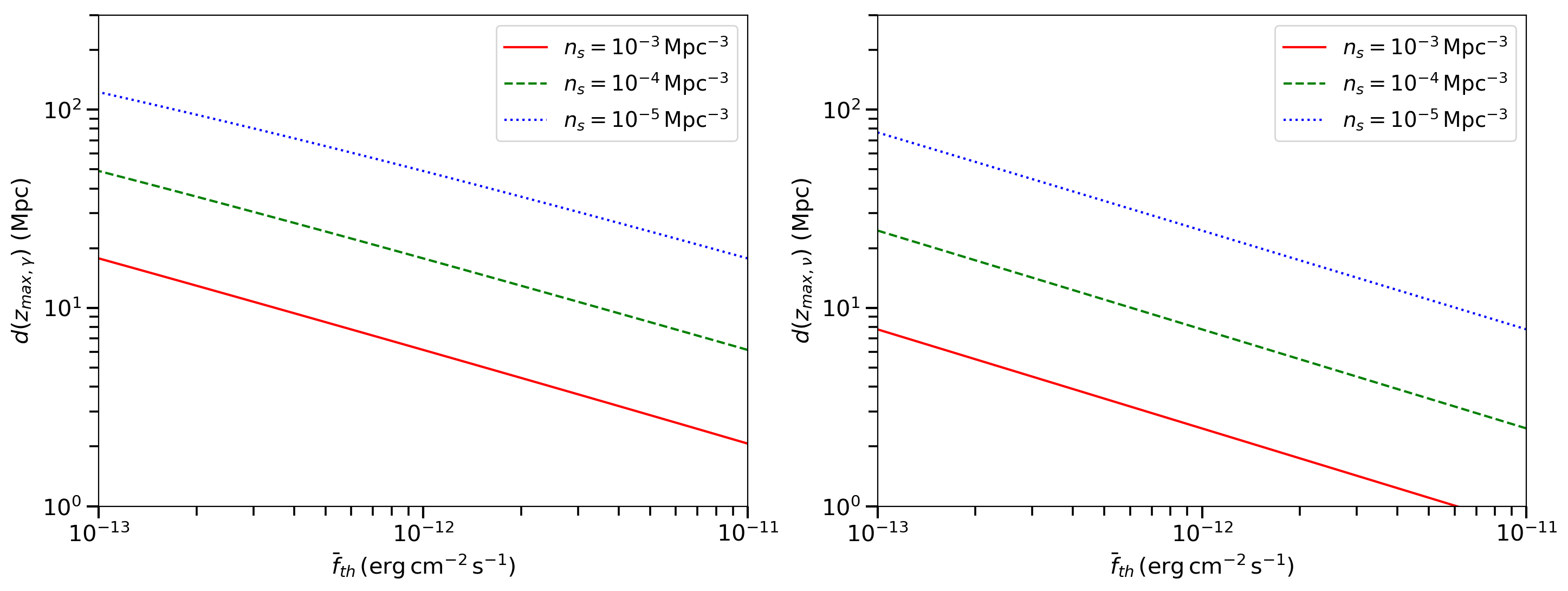}
    \caption{The horizon distance as a function of telescope sensitivities. Left panel: observation of \mbox{>2 TeV} gamma-rays. Right panel: observation of >100 TeV neutrinos. \xadded{The three lines from top to bottom correspond to three source number densities, $10^{-3}$, $10^{-4}$, and~$10^{-5} \, \rm Mpc^{-3}$.}}
    \label{fig1}
\end{figure}

Based on the method and assumptions above, we calculate the horizon distance and the source detection probability. We consider observations of gamma-rays above 2~TeV % and use sensitivity around WCDA level \citep{WCDA_sensitivity}, 
with telescope sensitivity in the range of about $10^{-13}$--$10^{-11} \rm \, erg \, cm^{-2} \, s^{-1}$. 
For neutrino observations, the~sensitivity for neutrino source detection at >100 TeV is also about $10^{-12} \rm \, erg \, cm^{-2} \, s^{-1}$. The~horizons for gamma-ray and neutrino telescopes are shown in Figure~\ref{fig1} for various source density $n_s$.
%We choose three different value $10^{-4}$, $10^{-5}$, $10^{-6}$ $\rm Mpc^{-3}$ for source number density.
\textls[-11]{The gamma-ray horizon is roughly scaling with the sensitivity as $d(z_{\max})$$\propto$$\bar{f}_{th}^{-1/2}$. And~we find that the horizon is <100 Mpc for high energy gamma-ray sources. The~EBL absorption does not affect the horizon significantly, and~the limitation on the gamma-ray source detection is mainly determined by the telescope~sensitivity}.

For the UHECR production rate and spectrum, %$\alpha$ is the power law index of UHECR spectrum in the source region and $m$ describe the redshift evolution of source number density, 
we take $\alpha = 2$ for particle acceleration in collisionless shocks \citep{shock_acc}, and~$m = 3$, comparable to star formation rate evolution and active galactic nuclei redshift distribution \citep{QSO_z_evo}. As~for the effective area of the UHECR array we assume the energy dependence following the Auger experiment \citep{Aeff_Auger}, i.e.,~beyond  $10^{18.5} \rm \, eV$, the~detection reaches a full efficiency and $A_{eff}$ is a constant independence of $E$.

%is adopted to be $A_{eff} \propto lg(E/eV)-17.5$ when UHECRs' energy $< 10^{18.5} \rm \, eV$. For UHECRs above $10^{18.5} \rm \, eV$, we assume array reach full efficiency and $A_{eff}$ have a stable value. This form of effective area is a simplification of Auger's trigger efficiency \citep{Aeff_Auger}

\begin{figure}[H]
    \includegraphics[width=0.98\textwidth]{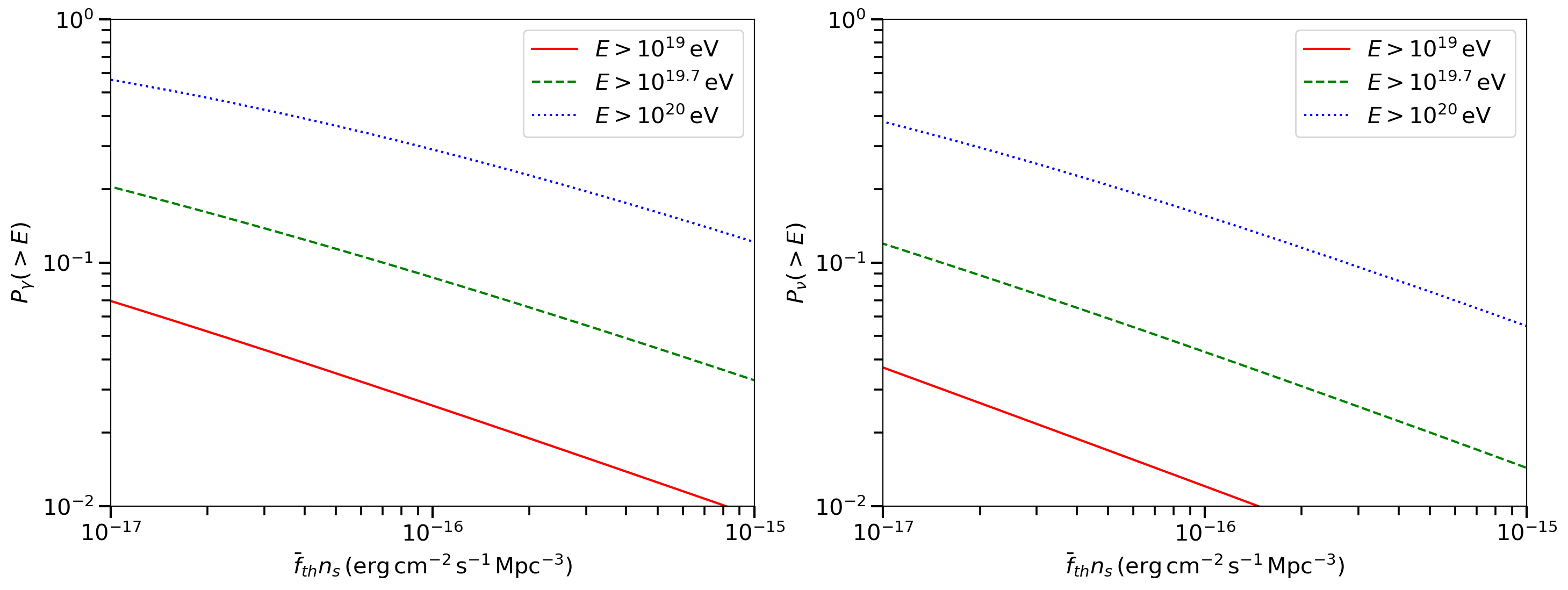}
    \caption{Average detection probability for gamma-ray/neutrino sources toward UHECR directions as a function of the product of source number density and the gamma-ray/neutrino telescope sensitivity. Left panel: observation of >2 TeV gamma-rays.  Right panel: observation of >100 TeV neutrinos. The~three lines correspond to three lower limits of cosmic ray energy, $E > 10^{19} \, \rm eV$, $E > 10^{19.7} \, \rm eV$ and $E > 10^{20} \, \rm eV$.
    %solid line, dashed line and dot line for source number density $10^{-3} \, \rm Mpc^{-3}$, $10^{-4} \, \rm Mpc^{-3}$ and $10^{-5} \, \rm Mpc^{-3}$. 
   }
    \label{fig2}
\end{figure}

With the horizons derived, the~probability of source detection per UHECR can be calculated. Three parameters determine the detection probability: the sensitivity, source number density and the lower limit of cosmic ray energy. %lower limit of UHECRs sample can affect whole number of UHECRs we use. These are three main physical parameters affect $P(>E)$. 
The first two parameters affect the horizon together by their product, $\bar{f}_{th} n_{s}$, because~the maximum redshift $z_{max}$ can be seperated from $n_s$ and $\bar{f}_{th}$ in Equations~(3) and (4).
The dependence of the detection probability on the product, $\bar{f}_{th} n_{s}$, is shown in Figure~\ref{fig2}. A~larger source number density leads to lower luminosity, and~then smaller detection probability. A~better gamma-ray or neutrino telescope sensitivity surely leads to a higher~probability.
\vspace{-3pt}

%\begin{figure}[!ht]
%    \centering
%    \includegraphics[width=\textwidth]{detect_probability_density.png}
%    \caption{Detection probability for gamma ray an neutrino source at UHECR %direction versus source number density. Left panel: detection probability for %gamma-ray source at UHECR direction, red and blue color stand for UHECR energy %$\> 10^{19} \, \rm eV$ and $\> 10^{20} \, \rm eV$, solid line, dashed line and %dot line for sensitivity $10^{-11}$, $10^{-12}$ and $10^{-13} \, \rm erg\, %cm^{-2} \, s^{-1}$. Right panel: detection probability for neutrino source at %UHECR direction, the meaning of linestyle and color is the same as that for left panel}
%    \label{fig:my_label}
%\end{figure}

%In Figure~2 we show the prediction especially for LHAASO/WCDA and IceCube.
Prediction for LHAASO-WCDA and IceCube can be obtained from Figure~\ref{fig2}.
For the WCDA sensitivity of $\bar{f}_{th} = 10^{-12} \rm \, erg \, cm^{-2} \, s^{-1}$\citep{LHAASO_sciencebook} and assuming a source number density of $n_s = 10^{-4} \rm \, Mpc^{-3}$, the~detection probability of gamma-ray sources at the directions of UHECRs with energy above $10^{20}\, \rm eV$ can be quite high, $P_{\gamma}$(>10$^{20 \, \rm eV})$$\sim$$10\%$. Similar probability is obtained for neutrino source searching with IceCube, which has a typical sensitivity of $\bar{f}_{th} = 10^{-9} \rm \, GeV\, cm^{-2}\, s^{-1}$ \citep{IceCube_sensitivity}.

\begin{figure}[H]
    \includegraphics[width=\textwidth]{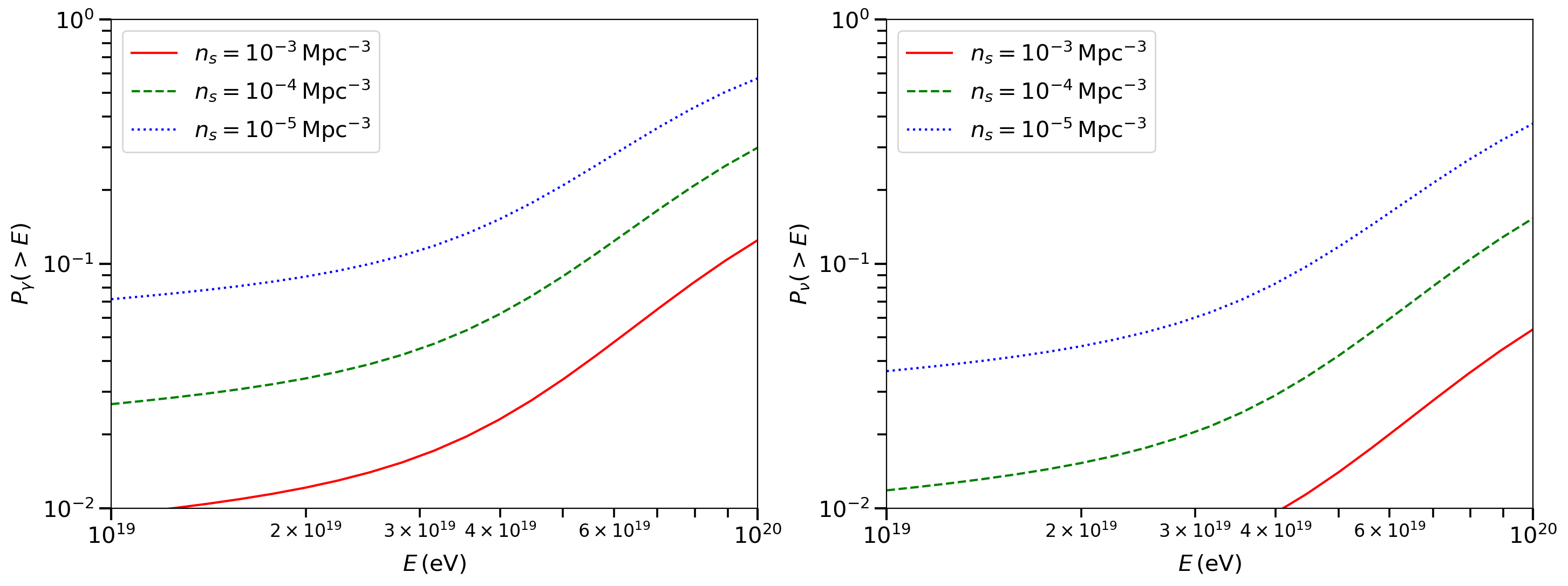}
    \caption{Average detection probability for gamma-ray/neutrino sources toward UHECR directions as a function of the lower limit of the cosmic ray energy. Left panel: observation of >2 TeV gamma-rays.  Right panel: observation of >100 TeV neutrinos. The~three lines correspond to three source number densities, $10^{-3}$, $10^{-4}$, and~$10^{-5} \, \rm Mpc^{-3}$. The~sensitivity is assumed to be $\bar{f}_{th} = 10^{-12} \rm \, erg \, cm^{-2} \, s^{-1}$ for both gamma-ray and neutrino~telescopes.}
    \label{fig3}
\end{figure}

In Figure~\ref{fig3} we show the dependence of the detection probability on the energy lower limit of UHECRs sample, assuming $\bar{f}_{th}=10^{-12}\rm erg\,cm^{-2}s^{-1}$. The~probability increases with cosmic ray energy. For~higher energy cosmic rays, the~GZK effect will suppress UHECR flux from distant sources, leading to a higher detection probability for gamma-ray or neutrino sources. Above~energy $10^{19.5} \rm \, eV$, GZK cutoff suppresses the observed UHECR flux more significantly, and~lead to a fast increasing of detection~probability.

\unskip

%%%%%%%%%%%%%%%%%%%%%%%%%%%%%%%%%%%%%%%%%%
\section{Conclusions and~Discussion}

In this work, we show a picture for UHECR source bearing galaxies as high energy gamma-ray and neutrino sources, motivated by the coincidence of IceCube neutrino flux and Waxman-Bahcall bound. We estimate the probability of detecting gamma-ray or neutrino sources around UHECR directions. The~detection probability for different parameters, including the lower limit of UHECR energy, the~source number density and the sensitivity of gamma-ray and neutrino telescopes, is estimated. We find that the average probability per UHECR is about $10\%$ for detecting gamma-ray sources at the directions of UHECRs with energy $\gtrsim 10^{19} \rm \, eV$, for~a gamma-ray telescope with a sensitivity of $\bar{f}_{th} = 10^{-12} \rm \, erg \, cm^{-2} \, s^{-1}$ and if the source density number density is $n_s=10^{-5} \rm \, Mpc^{-3}$. This implies that it needs a UHECR sample containing about 10 random events of $E > 10^{20} \rm \, eV$ to find a gamma-ray source. The~probability for neutrino source detection is found to be in the same order of $10\%$ for the same parameter setting, because~the gamma-ray and neutrino emission is well connected in the flux and~spectrum.

The model can be used for both steady sources and transient sources of UHECRs. For~steady sources, the~source number density in our model and calculation just means their number density. For~example, starbusrt galaxies are one kind of potential sources for UHECRs \citep{SBG_UHECR_hotspot}. Number density of starburst galaxies is about $10^{-5} \rm \, Mpc^{-3}$, and~they can be seen as steady source. When the gamma-ray telescope sensitivity is $10^{-12} \rm \, erg \, cm^{-2} \, s^{-1}$, we have $10\%$ probability to find gamma-ray source at direction of a UHECR with \mbox{energy $\gtrsim 10^{19} \rm \, eV$}, if~source number density is at a level of that for starburst galaxies.
For transient sources, gamma-ray and neutrino emission from their host galaxies still last $\Delta t_{\pi}$$\sim$$10^{7} \rm \, yr$. The~production, $\dot{n}_s \Delta t_{\pi}$, is an average number density for transient sources. But~these transient events often occur repeatedly in their host galaxies, and~we consider the host galaxies as the gamma-ray and neutrino sources. To~get the number density of host galaxies, the~average number density of the sources should multiply a factor $\Delta t_{int} / \Delta t_{\pi}$, where $\Delta t_{int}$ is the typical time interval between two transient events. So we can use $n_s$$\sim$$\dot{n}_{s} \Delta t_{\pi}$ $(\Delta t_{int} / \Delta t_{\pi})$ as the source number density for transient sources. Gamma-ray bursts (GRBs) are candidate sources for UHECRs \citep{Waxman_UHECRGRB,Kohta_UHECRGRB,Vietri_GRB}, and~they are transient events. Event rate of GRB is about $\dot{n}_{GRB}$$\sim$$(4\pi/\Delta \Omega) \times 1\rm \, Gpc^{-3} \, yr^{-1}$$\sim$$10^{-7}\rm \, Mpc^{-3} \, yr^{-1}$, if~we consider the beaming effect. When we take the time interval $\Delta t_{int}$$\sim$$10^{3}$ $ \mathrm{yr}$, we get source number density for GRBs is $n_{s,\mathrm{GRB}}$$\sim$$10^{-4} \rm \, Mpc^{-3}$. For~source number density at this level, an~UHECR sample with energy above $10^{20} \rm \, eV$ is needed to get a $10\%$ detection~probability.

\xadded{We consider protons as the main composition of UHECRs. A~heavier composition will affect our results. For~heavy nuclei, the~Lamour radius will be smaller, $R_{L} = E/ZeB$, which lead to a stronger deflection. The~typical deflection angle for heavy nuclei is \emph{Z} times as large as that for proton, $\theta_{Z} = Z \theta_{\rm proton}$. It is hard to find gamma-rays or neutrino sources in a small region around UHECR direction when the UHECRs composition is too heavy. And~if the search is in a larger region, there will be more noise and false positive signals.} 

When we calculate the typical gamma-ray luminosity of single source in Section~\ref{sec3.1}, we consider the high energy gamma-rays all from $\pi^{0}$ decay. Besides~the gamma-rays hadronic process, there are other leptonic processes, Inverse-Compton and synchrotron radiation of electrons, producing high energy gamma-rays. We assume gamma-rays from $\pi^{0}$ decay typically account for a fraction $f$ ($f \leq 1$) of the total gamma-ray luminosity, the~calculation in Sections~\ref{sec3} and \ref{sec4} just takes $f = 1$. When leptonic components contribute significantly, the~gamma-ray luminosity should be modified by multiplying a factor \xadded{of} $1/f$. The~effect of this modification is the same as multiplying a factor $f$ on source number density or on telescope~sensitivity.

%%%%%%%%%%%%%%%%%%%%%%%%%%%%%%%%%%%%%%%%%%
%\section{Patents}

%This section is not mandatory, but may be added if there are patents resulting from the work reported in this manuscript.

%%%%%%%%%%%%%%%%%%%%%%%%%%%%%%%%%%%%%%%%%%
\vspace{6pt} 

%%%%%%%%%%%%%%%%%%%%%%%%%%%%%%%%%%%%%%%%%%
%% optional
%\supplementary{The following supporting information can be downloaded at:  \linksupplementary{s1}, Figure S1: title; Table S1: title; Video S1: title.}

% Only for the journal Methods and Protocols:
% If you wish to submit a video article, please do so with any other supplementary material.
% \supplementary{The following supporting information can be downloaded at: \linksupplementary{s1}, Figure S1: title; Table S1: title; Video S1: title. A supporting video article is available at doi: link.}

%%%%%%%%%%%%%%%%%%%%%%%%%%%%%%%%%%%%%%%%%%
\authorcontributions{Conceptualization, Z.L.; methodology, Z.L. and Q.Z.; writing---original draft preparation, Q.Z.; writing---review and editing, Z.L., X.T. and Q.Z.; supervision, Z.L.; project administration, Z.L.; funding acquisition, Z.L. All authors have read and agreed to the published version of the~manuscript.}

\funding{This work is supported by the Natural Science Foundation of China (No. 11773003, U1931201) and the China Manned Space Project (CMS-CSST-2021-B11).}

\dataavailability{Not applicable %MDPI: In this section, please provide details regarding where data supporting reported results can be found, including links to publicly archived datasets analyzed or generated during the study. Please refer to suggested Data Availability Statements in section ``MDPI Research Data Policies'' at \url{https://www.mdpi.com/ethics}. If the study did not report any data, you might add ``Not applicable'' here.
}

\acknowledgments{The authors thank Tianqi Huang and Kai Wang for helpful~discussions.}

\conflictsofinterest{The authors declare no conflict of~interest.} 

\begin{adjustwidth}{-\extralength}{0cm}
%\printendnotes[custom] % Un-comment to print a list of endnotes

\reftitle{References}

\end{adjustwidth}
\end{document}